\documentclass[a4paper,twocolumn,english,prl,superscriptaddress,showpacs,longbibliography]{revtex4-1}
\usepackage[latin9]{inputenc}
\usepackage{amsmath}
\usepackage{amssymb}
\usepackage{graphicx}

\makeatletter

\pdfpageheight\paperheight
\pdfpagewidth\paperwidth

\usepackage[colorlinks,citecolor=darkblue,linkcolor=darkred,urlcolor=darkblue] {hyperref}

\usepackage{babel}
\usepackage{xcolor}
\usepackage{algorithmicx}
\usepackage{algpseudocode}

\usepackage[backgroundcolor=green!50!white]{todonotes}
\definecolor{darkblue}{rgb}{0.1,0.2,0.6} 
\definecolor{darkred}{rgb}{0.8,0.1,0.2}
\usepackage[all]{hypcap}

\renewcommand{\BibitemShut}[1]{}

\makeatother

\usepackage{babel}
\begin{document}
\global\long\def\E{\mathrm{e}}
\global\long\def\D{\mathrm{d}}
\global\long\def\I{\mathrm{i}}
\global\long\def\mat#1{\mathsf{#1}}
\global\long\def\cf{\textit{cf.}}
\global\long\def\ie{\textit{i.e.}}
\global\long\def\eg{\textit{e.g.}}
\global\long\def\vs{\textit{vs.}}
 \global\long\def\ket#1{\left|#1\right\rangle }

\global\long\def\etal{\textit{et al.}}
\global\long\def\tr{\text{Tr}\,}
 \global\long\def\im{\text{Im}\,}
 \global\long\def\re{\text{Re}\,}
 \global\long\def\bra#1{\left\langle #1\right|}
 \global\long\def\braket#1#2{\left.\left\langle #1\right|#2\right\rangle }
 \global\long\def\obracket#1#2#3{\left\langle #1\right|#2\left|#3\right\rangle }
 \global\long\def\proj#1#2{\left.\left.\left|#1\right\rangle \right\langle #2\right|}

\title{Information propagation in isolated quantum systems}

\author{David J. Luitz}

\affiliation{Department of Physics and Institute for Condensed Matter Theory,
University of Illinois at Urbana-Champaign, Urbana, Illinois 61801,
USA}

\affiliation{Department of Physics, T42, Technische Universität München, James-Franck-Straße
1, D-85748 Garching, Germany}
\email{david.luitz@tum.de}

\author{Yevgeny Bar Lev}

\affiliation{Department of Chemistry, Columbia University, 3000 Broadway, New
York, New York 10027, USA}
\email{yevgeny.barlev@columbia.edu}

\begin{abstract}
Entanglement growth and out-of-time-order correlators (OTOC) are used
to assess the propagation of information in isolated quantum systems.
In this work, using large scale exact time-evolution we show that
for weakly disordered nonintegrable systems information propagates
behind a ballistically moving front, and the entanglement entropy
growths linearly in time. For stronger disorder the motion of the
information front is algebraic and sub-ballistic and is characterized
by an exponent which depends on the strength of the disorder, similarly
to the sublinear growth of the entanglement entropy. We show that
the dynamical exponent associated with the information front coincides
with the exponent of the growth of the entanglement entropy for both
weak and strong disorder. We also demonstrate that the temporal dependence
of the OTOC is characterized by a fast \emph{nonexponential} growth,
followed by a slow saturation after the passage of the information
front. Finally,we discuss the implications of this behavioral change
on the growth of the entanglement entropy.
\end{abstract}
\maketitle
\emph{Introduction.} \textemdash{} While the speed of light is the
absolute upper limit of information propagation in both classical
and quantum relativistic systems, surprisingly a velocity which plays
a similar role exists also for short-range interacting \emph{nonrelativistic}
quantum systems. This velocity, known as the Lieb-Robinson velocity,
bounds the spreading of correlations in the system and implies that
information about local initial excitations propagates within a causal
``light-cone'', similarly to the light-cone encountered in the theory
of special relativity \cite{Lieb1972}. The shape of the light-cone
can be obtained from the correlation function
\begin{equation}
C_{i}^{\beta}\left(t\right)=-\frac{1}{Z}\tr\E^{-\beta\hat{H}}\left[\hat{A}_{i}\left(t\right),\hat{B}_{j=0}\right]^{2},
\end{equation}
where $\hat{H}$ is the Hamiltonian, $\beta$ is the inverse temperature,
$\hat{A}_{i}(t)$ and $\hat{B}_{j=0}$ are local Hermitian operators
in the Heisenberg picture operating on sites $i$ and $j=0$ , $\left[.,.\right]$
is the commutator and $Z=\tr\E^{-\beta\hat{H}}$ is the partition
function. Lieb and Robinson proved that for short-range interacting
Hamiltonians this correlation function, commonly know as the out-of-time-order
correlator (OTOC), is bounded by $C_{i}^{\beta}\left(t\right)\leq c\exp\left[-a\left(i-vt\right)\right],$
where $a,\,c$ are constants and $v$ is the Lieb-Robinson (LR) velocity.
The OTOC was first introduced by Larkin and Ovchinnikov \cite{Larkin1969},
who noted that it embodies a signature of classical chaos in the corresponding
quantum system. In the semiclassical limit the commutator is replaced
by Poisson brackets and for the choice of the operators $\hat{A}\left(t\right)\to q\left(t\right)$
and $\hat{B}\to p$, gives $C^{\beta}\left(t\right)\sim\hbar^{2}\left(\partial q\left(t\right)/\partial q\right)^{2}$.
The OTOC therefore measures the sensitivity of classical trajectories
to their initial conditions, which for chaotic systems implies that
the OTOC grows exponentially in time, $C^{\beta}\left(t\right)\sim\exp\left[2\lambda_{L}t\right]$,
where $\lambda_{L}$ is the classical Lyapunov exponent. 
\begin{figure}[t]
\includegraphics[width=1\columnwidth]{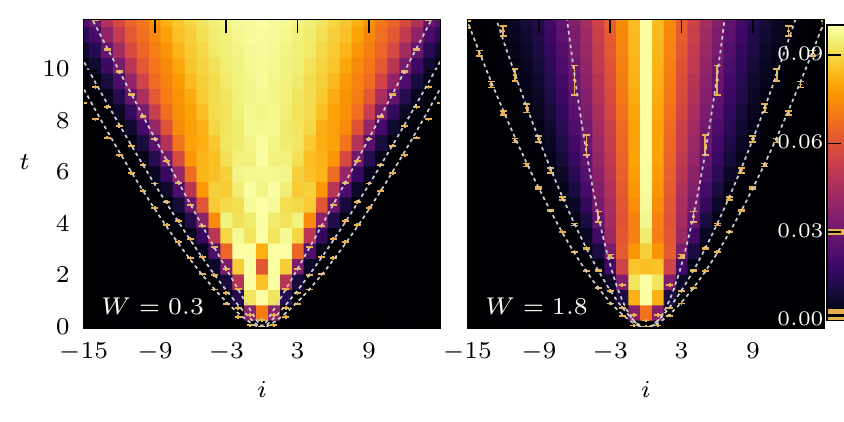}

\caption{\label{fig:L31OTOC}OTOC {[}Eq. (\ref{eq:otoc}){]} for the $L=31$
random Heisenberg chain in the $S_{z}=\frac{1}{2}$ sector at weak
disorder $W=0.3$ (left) and intermediate disorder $W=1.8$ (right).
At weak disorder, a linear light cone is visible (illustrated by contour
lines at three thresholds indicated on the colorbar), which changes
to a power-law light cone at intermediate disorder, with considerably
slower information spreading. Here, we average over only a small number
of disorder realizations ($n=10$ for $W=0.3$ and $n=45$ for $W=1.8$),
but we symmetrize the OTOC, effectively doubling the number of realizations. }
\end{figure}
A related manifestation of classical chaos in quantum systems was
introduced in Ref.~\cite{Jalabert2001a}. The time-scale $t_{d}\sim\lambda_{L}^{-1}$
is a purely classical time-scale, and quantum effects become appreciable
only on a parametrically longer time-scale known as the Ehrenfest
time, $t_{\text{Ehrenfest}}\sim\lambda_{L}^{-1}\ln\left[s_{cl}/\hbar\right]$,
with the typical classical action $s_{cl}$ \cite{Berman1978,Chirikov1981,Chirikov1988,Aleiner1997}.
For quantum systems without a proper semiclassical limit the concept
of Ehrenfest time does not directly apply, and the exponential growth
of the OTOC, while plausible, cannot be similarly motivated. Moreover,
for quantum systems with a finite local Hilbert space dimension it
is easy to show that the OTOC is bounded from above uniformly in time.
Indeed the growth saturates after a time known as the ``scrambling
time,'' $t_{sc}$ \cite{sekino_fast_2008,Maldacena2016a}. This creates
an additional hurdle for the observation of the exponential growth
of the OTOC in systems with a small local Hilbert space dimension,
since the regime of exponential growth is pronounced only for times
$t_{d}\ll t\ll t_{sc}$. This difficulty can be remedied either by
increasing the local Hilbert space dimension, or by studying OTOCs
of extensive operators as recently proposed in Ref.~\cite{Kukuljan2017}.

The interest in OTOCs was revived by Kitaev \cite{Kitaev2015,Kitaev2015a},
who using the AdS-CFT correspondence established a duality between
some strongly coupled quantum systems and black-holes \cite{Georges2000,Georges2001,Maldacena2016,Maldacena2016a}.

The spreading of the OTOCs in space directly corresponds to the spreading
of information on local excitations. Surprisingly, while transport
in generic systems is diffusive, the LR bound suggests that information
spreads behind a ballistically propagating front, namely that it resides
within a linear light-cone. This was observed in the study of the
growth of the entanglement entropy, a global measure of quantum information,
where it was conjectured that entanglement is transmitted ``on contact,''
similarly to the spread of fire, and therefore inherently spreads
faster than particles or energy \cite{Kim2013}. The ballistic spreading
of OTOCs was directly established and linked to combustion theory
in Ref.~\cite{Aleiner2016}. In this work a relation between the
entanglement entropy growth and the spreading of the OTOC was also
conjectured \cite{Aleiner2016} (c.f. Ref.~\cite{Fan2016} for a
connection to the second Rényi entropy).

In this work we examine the spreading of quantum information using
both OTOCs and the entanglement entropy (EE) growth and establish
the relationship between the two for diffusive, and subdiffusive systems.
We also provide a detailed analysis of the shape of the OTOCs in space
and time.

\emph{Model.} \textemdash{} We study the one-dimensional spin\textendash $\frac{1}{2}$
Heisenberg chain of length $L$ in a random magnetic field,

\begin{equation}
\hat{H}=J\sum_{i=1}^{L-1}\vec{S}_{i}\cdot\vec{S}_{i+1}+\sum_{i=1}^{L}h_{i}\hat{S}_{i}^{z},\label{eq:disorder_Heisenberg}
\end{equation}
 with the coupling between the spins $J=1$ , and the random fields
$h_{i}\in\left[-W,W\right]$ drawn from a uniform distribution with
disorder strength $W$. This model exhibits an ergodic to nonergodic
transition \cite{Basko2006a}, which for infinite temperature occurs
at $W_{c}\approx3.7$ \cite{Berkelbach2010a,Luitz2015}. Interestingly
even in the nonergodic phase, where transport is completely frozen,
information continues to spread logarithmically in time via dephasing,
as was initially established using the growth of the EE \cite{Znidaric2008,Bardarson2012}
and later using OTOCs \cite{Fan2016,He2016,Chen2016,chen_out--time-order_2016,Swingle2016,Slagle2016,Deng2016a}.
In contrast, the spreading of information in noninteracting Anderson
insulators is completely frozen \cite{Hamza2012,Friesdorf}. The ergodic
phase of this model, which occurs for $W<W_{c}$, exhibits anomalous
subdiffusive spin transport characterized by a dynamical exponent
varying continuously with disorder strength \cite{Lev2014,Agarwal2014,Luitz2015a,Gopalakrishnan2015a,Luitz2016,Lerose2015,Znidaric2016,Khait2016,Luitz2016b},
but also a sublinear EE growth \cite{Luitz2015a} (cf. Ref. \cite{luitz_ergodic_2016}
for a review). Here we focus solely on the ergodic phase and on the
infinite temperature limit. In the weakly disordered limit, the disorder
can be considered as an integrability breaking perturbation, allowing
us to draw conclusions on generic \emph{clean} systems. In this limit,
information spread is bounded by normal linear light cones, as displayed
in the left panel of Fig.~\ref{fig:L31OTOC}. For stronger disorder,
information is contained within \emph{anomalously shaped} light-cones,
which are well described by power-laws (see right panel of Fig.~\ref{fig:L31OTOC}).
Such anomalous light-cones were previously predicted to exist for
the $XY$ spin chain in a quasiperiodic potential \cite{Damanik2014},
however to the best of our knowledge were never observed.

\emph{Calculation of the OTOC.} \textemdash{} For the system (\ref{eq:disorder_Heisenberg}),
the OTOC can be simplified to
\begin{equation}
C_{i}^{\beta=0}\left(t\right)=\frac{1}{8}-\frac{2}{Z}\text{Re }\tr\,\left(\hat{S}_{i}^{z}\left(t\right)\hat{S}_{j=0}^{z}\hat{S}_{i}^{z}\left(t\right)\hat{S}_{j=0}^{z}\right),\label{eq:otoc}
\end{equation}
where to utilize the conservation of the total spin we take, $\hat{A}_{i}\left(t\right)=\hat{S}_{i}^{z}\left(t\right)$
and $\hat{B}_{j=0}=\hat{S}_{j=0}^{z}$. The numerical calculation
of the correlation function in (\ref{eq:otoc}) for large enough system
sizes is a challenging task and several approaches have been used
previously, relying on the propagation of operators in the Heisenberg
picture using either an exact diagonalization (ED) of the Hamiltonian
\cite{chen_out--time-order_2016} or a representation in terms of
matrix product operators (MPO) \cite{Bohrdt2016}. Although these
approaches yield accurate results, they severely suffer from an exponential
scaling with either system size (ED) or time (MPO). In order to alleviate
these problems we calculate the OTOC in the Schrödinger picture using
an exact time-evolution with a Krylov space method \cite{nauts_new_1983,hernandez_slepc:_2005,luitz_ergodic_2016}.
Our method still scales exponentially in system size, however much
larger system sizes can be reached compared to exact diagonalization
of the Hamiltonian. In this work we report results for system sizes
up to $L=31$ (cf. Fig. \ref{fig:L31OTOC}). To evaluate the trace
in (\ref{eq:otoc}) we utilize Lévy's lemma and the concept of quantum
typicality \cite{popescu_entanglement_2006,goldstein_canonical_2006,reimann_typicality_2007,luitz_ergodic_2016}.
The trace is approximated by an expectation value with respect to
a pure random state $\ket{\tilde{\psi}}$ drawn from the Haar measure,
such that, $C_{i}\left(t\right)\approx\left\langle \tilde{\psi}\left|\hat{S}_{i}^{z}\left(t\right)\hat{S}_{0}^{z}\hat{S}_{i}^{z}\left(t\right)\hat{S}_{0}^{z}\right|\tilde{\psi}\right\rangle $.
As was shown by Lévy the error of this approximation is inversely
proportional to the dimension of the Hilbert space, namely it is exponentially
small in the length of the lattice, $L$. It is convenient to calculate
independently $\ket{\psi_{1}\left(t\right)}=\hat{S}_{i}^{z}\left(t\right)\hat{S}_{0}^{z}\ket{\tilde{\psi}}$
and $\ket{\psi_{2}\left(t\right)}=\hat{S}_{0}^{z}\hat{S}_{i}^{z}\left(t\right)\ket{\tilde{\psi}}$,
for various values of $i$. These calculations employ the exact propagation
of a single wavefunction using a projection of the matrix exponential
$\exp\left[-\I Ht\right]$ on the Krylov space of the Hamiltonian
$H$ (cf. Ref.~\cite{luitz_ergodic_2016} Sec V.A. for details).
The OTOC is then given by the overlap $C_{i}\left(t\right)=\braket{\psi_{2}\left(t\right)}{\psi_{1}\left(t\right)}$.
While at each time step a full propagation back to time $t=0$ is
necessary, this procedure can be carried out efficiently, allowing
us to access system sizes up to $L=31$ (Hilbert space dimension $3\cdot10^{8}$).

\emph{Tomography of the OTOC.} \textemdash{} 
\begin{figure}
\includegraphics{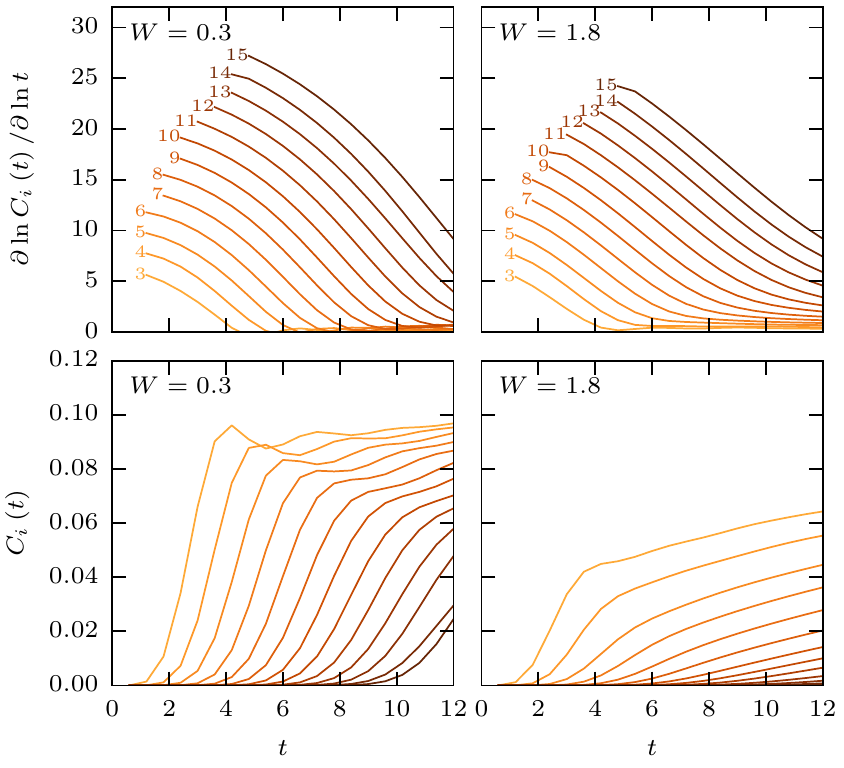}

\caption{\label{fig:L31fixedSpace}OTOC at fixed distances (indicated by numbers)
from the initial excitation in the middle of the lattice as function
of time, for two disorder strengths $W=0.3$ (bottom left) and $1.8$
(bottom right). Darker colors represent longer distances, $x=3-15$.
Upper panels show the logarithmic derivative of the corresponding
bottom panel. System size is $L=31$.}
\end{figure}
\begin{figure}
\includegraphics{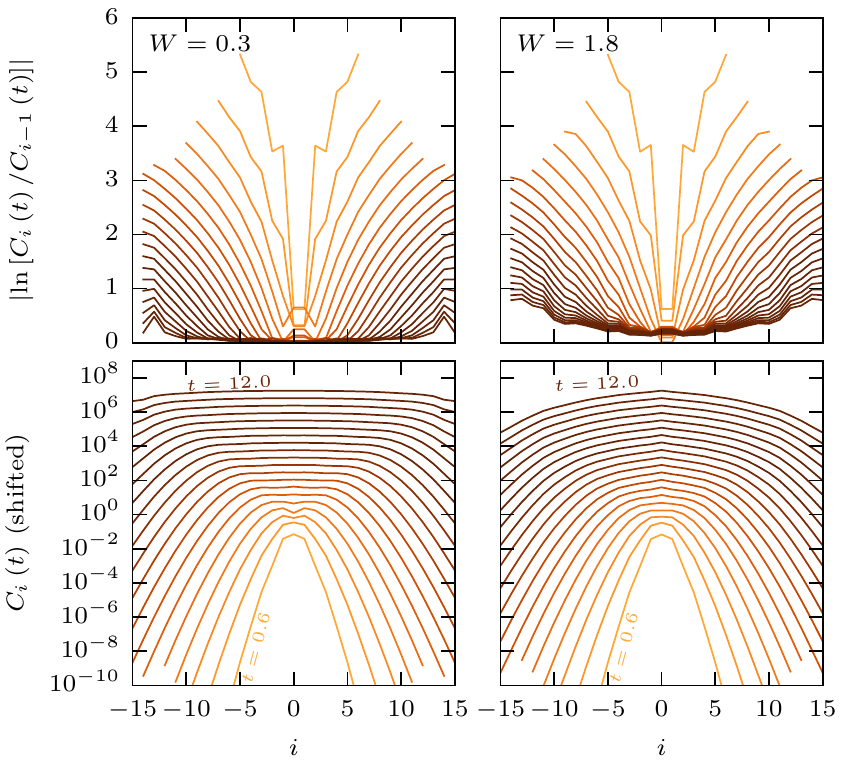}

\caption{\label{fig:L31fixedTime}OTOC at fixed times as function of site index,
for two disorder strengths $W=0.3$ (bottom left) and $1.8$ (bottom
right). Darker colors represent later times on a linear grid, $t=0.6\dots12$,
and the lines are shifted for clarity. Upper panels show the semi-logarithmic
derivative of the corresponding bottom panel. System size is $L=31$.}
\end{figure}
As explained in the introduction, the OTOC for two local operators
with a fixed distance is expected to grow exponentially with time.
In Fig.~\ref{fig:L31fixedSpace} we test this assertion for a number
of distances and two disorder strengths. The growth of the OTOC shows
two clear regimes: a fast initial growth, associated with the advance
of the information front, followed by a slow saturation to the maximal
value after the front has already passed (bottom panels of Fig.~\ref{fig:L31fixedSpace}).
Surprisingly the growth is not well described by an exponential, even
for very weak disorder ($W=0.3$), as is apparent from the logarithmic
derivative on the upper panel of Fig.~\ref{fig:L31fixedSpace}. This
derivative monotonically decreases to zero, \emph{without} a visible
reversal trend for larger distances and longer times. For stronger
disorder, within the subdiffusive phase, one might suspect either
a stretched exponential \cite{Damanik2014} or a power law growth
of the OTOC \cite{Fan2016}. However, our data do not support these
forms, possibly due to the very limited time range of their validity.

In Fig.~\ref{fig:L31fixedTime} we study the spatial profile of the
OTOC for fixed times. The LR bound establishes that the spatial decay
of the OTOC should be \emph{at} \emph{least} exponential. In Fig.~\ref{fig:L31fixedTime}
we show the spatial profiles for different times (lower panels) and
the corresponding semi-logarithmic derivative in space (upper panels).
The decay appears to be faster than exponential, suggesting that the
LR bound is not saturated, however for longer times and distances
the profile does appear to converge to an exponential form, which
is more apparent for the case of stronger disorder.

\emph{Light cone shape.} \textemdash{} 
\begin{figure}
\includegraphics{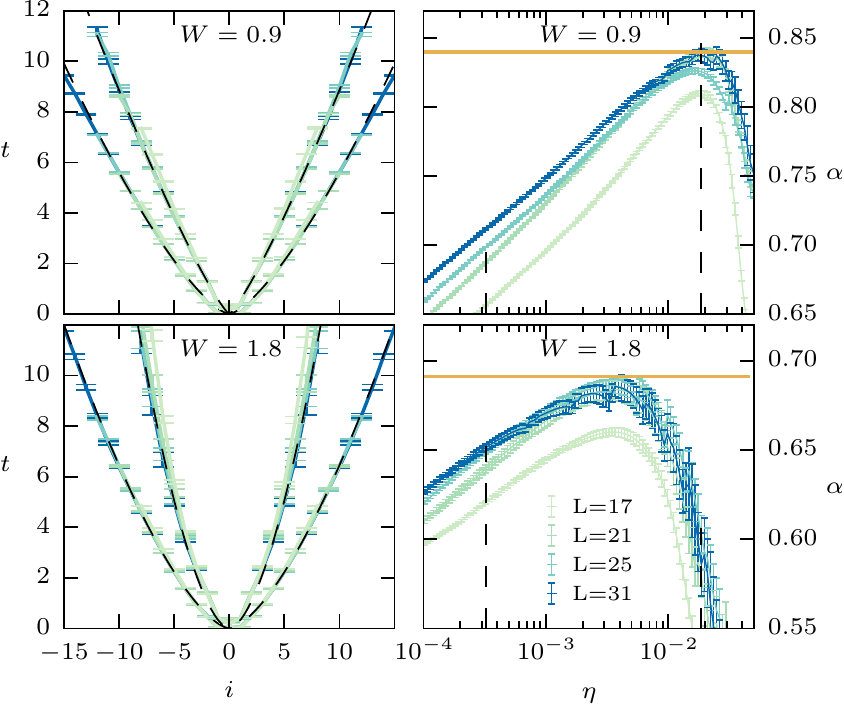}

\caption{\label{fig:size_comp}The left column illustrates the extraction of
the dynamical exponents from the shape of the OTOC ``light-cones''
for two disorder strengths $W=0.9$ and $1.8$, two thresholds (the
two distinctive groups of colored lines on each panel) and various
system size, $L=17,\,21,\,25,$ and $31$ (larger sizes correspond
to more intense color). The dashed black lines are power law fits
to the contour lines. The dependence of the extracted dynamical exponent
on the threshold is plotted on the right column, for same disorder
strengths and system sizes. The dashed black line here mark the thresholds
used for the data on the left column, and the orange solid line represents
the final selection of the dynamical exponent, which does not depend
on the size of the system for large enough systems. Error bars on
both columns represent the statistical errors in the extraction of
the contours or the exponents, correspondingly.}
\end{figure}
\begin{figure}
\includegraphics{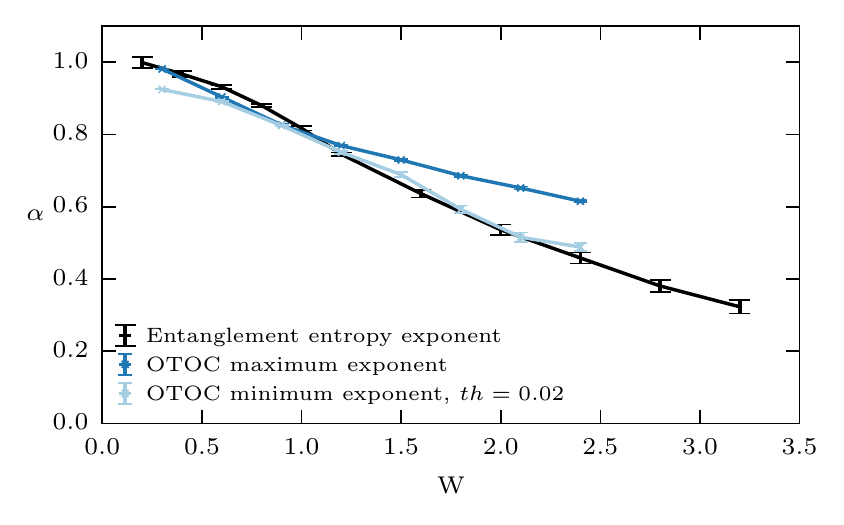}\caption{\label{fig:ExpComparison}Dynamical exponent relating space and time
as extracted from the shape of the OTOC ``light-cones'' for a system
size of $L=25$ (blue line), and the spread of the entanglement entropy
(orange line). The dynamical exponent for the entanglement entropy
was taken from Ref.~\cite{luitz_extended_2016} for $L=28$. The
error bars indicate the statistical uncertainty and do not include
systematic errors. As the exponent depends on the choice of the threshold,
we show the maximal exponent as well as the exponent of the front
propagation for large threshold ($\eta=0.02$).}
\end{figure}
A qualitative inspection of the OTOC in Fig.~\ref{fig:L31OTOC} reveals
a presumably linear propagation of the OTOC front at weak disorder
and a sublinear propagation at stronger disorder. To confirm this
observation more quantitatively, we extract the contour lines of the
OTOC, $C_{i}\left(t\right)=\eta$, as a function of space and time
for different thresholds, $10^{-4}<\eta<10^{-1}$ (the OTOC in (\ref{eq:otoc})
is bounded from above by 0.125). The contour lines are obtained from
the times at which the OTOC exceeds the threshold $\eta$ at each
lattice site, and the statistical error is estimated by a bootstrap
resampling. Motivated by the usual algebraic relation between space
and time, which applies for diffusive and subdiffusive systems, we
assume such a relation also for the shape of the contour lines, $x\sim t^{\alpha}$.
Typical results for different thresholds are shown in Figs.~\ref{fig:L31OTOC}
and \ref{fig:size_comp} with very good fits to this form. Since the
domain of the fits is clearly restricted in both space and time, we
proceed by assessing the finite size effects. The left panels of Fig.~\ref{fig:size_comp}
show that contour lines obtained for the same threshold $\eta$ but
different system sizes agree well within error bars. The right panels
of Fig.~\ref{fig:size_comp} show the exponent $\alpha$ as a function
of the threshold for various system sizes. Here we observe a strong
systematic dependence of the exponent on the value of the threshold,
such that $\alpha\left(\eta\right)$ exhibits a maximum at a threshold
of the order of $\eta\approx10^{-2}$. Finite size effects are the
strongest for very large and very small thresholds, where the power-law
fits are severely bounded either in space or in time. Hence we cannot
exclude that $\alpha\left(\eta\right)$ is a monotonic function in
the thermodynamic limit.

We interpret the maximal value of the dynamical exponent $\alpha$
as the \emph{fastest mode of information} \emph{spreading} in the
system, corresponding to the propagation of the tail of the OTOC,
and extract its value for different disorder strengths $W$ (see Fig.~\ref{fig:ExpComparison}
(blue)). The error bars in this figure are statistical and do not
include systematic errors, such as finite size effects. We observe
a monotonously decreasing exponent as a function of disorder strength,
starting from a value close to $\alpha=1$ at very weak disorder,
consistent with a linear light-cone. We compare this exponent to the
dynamical exponent obtained from the growth of the EE as a function
of time starting from a random product state (data is taken from Ref.~\cite{luitz_extended_2016}).
While the two exponents match very well at weak disorder, they seem
to deviate from each other starting from $W\approx1,$ suggesting
that the tail of the light cone spreads faster than EE. Extracting
the exponent $\alpha$ from contour lines obtained at a larger threshold
$\eta=0.02$ (or larger) \emph{does} produce a reasonable match, indicating
that the \emph{EE spreads as the front of the OTOC}.

\emph{Discussion.}\textbf{ }\textemdash{} We studied information spreading
in a generic quantum system using the OTOC. We showed that at fixed
distance, the temporal growth of the OTOC does \emph{not} appear to
have a finite regime of exponential growth neither for weak nor stronger
disorder, even for larger system sizes or longer distances. This suggests
that an exponential regime in \emph{local} quantum systems without
a semiclassical limit is either absent or very short. The spatial
profile of the OTOC seems to decay \emph{faster} than exponentially,
indicating that the LR bound could be further improved. However, we
note a weak trend towards an exponential profile at larger times and
stronger disorder.

We demonstrated that information mostly resides within spatio-temporal
light-cones. For weak disorder, with diffusive transport , we obtain
light-cones of linear shape. For stronger disorder, information transport
is suppressed, leading to a deformation of the linear light cone into
a power-law form, consistent with the previously observed subdiffusive
transport as well as with the sublinear algebraic growth of the EE
after a quench from a random product state. We directly compared the
dynamical exponents extracted from the tail and the propagating front
of the OTOC. While our data suggests that the tail of the OTOC propagates
\emph{faster} than the EE, the propagation of the front of the OTOC
and the EE coincide. Unlike the EE, the front of the OTOC thus provides
a glimpse into the local structure of information propagation in the
system.

We also demonstrated that the growth of the OTOC is markedly different
\emph{before} and \emph{after} its front passes through a given point
in space. A fast initial growth is followed by a much slower saturation
to the maximal value of the OTOC. This observation, combined with
the association between the EE growth and the propagation of the OTOC
front, which follows from our work, allows us to explain the apparent
slowing down of the initial fast growth of the EE, starting from a
product state, that was observed in a number of studies. We argue
that this slow saturation regime of both the EE and OTOC, occurs for
times $L^{1/\alpha}<t<t_{\text{Th}}\left(L\right)$, (where $t_{\text{Th}}$
is the generalized Thouless time, which scales algebraically with
system size \cite{Luitz2016b}), and is a consequence of the conservation
laws in the system. It is therefore expected to be absent for systems
without any conservation laws, such as certain Floquet systems. We
leave the study of information propagation in this regime for future
work.
\begin{acknowledgments}
YB would like to thank Igor Aleiner for many enlightening and helpful
discussions. This work was supported by National Science Foundation
Grant No. CHE-1464802 and by the Gordon and Betty Moore Foundation's
EPiQS Initiative through Grant No. GBMF4305 at the University of Illinois.
This research is part of the Blue Waters sustained-petascale computing
project, which is supported by the National Science Foundation (awards
OCI-0725070 and ACI-1238993) and the state of Illinois. Blue Waters
is a joint effort of the University of Illinois at Urbana-Champaign
and its National Center for Supercomputing Applications. Our code
uses the libraries PETSc \cite{petsc-efficient,petsc-user-ref,petsc-web-page}
and SLEPc \cite{hernandez_slepc:_2005}.
\end{acknowledgments}

\bibliographystyle{apsrev4-1}
\bibliography{lib_david,lib_yevgeny,lib_david_otoc}

\begin{thebibliography}{52}%
\makeatletter
\providecommand \@ifxundefined [1]{%
 \@ifx{#1\undefined}
}%
\providecommand \@ifnum [1]{%
 \ifnum #1\expandafter \@firstoftwo
 \else \expandafter \@secondoftwo
 \fi
}%
\providecommand \@ifx [1]{%
 \ifx #1\expandafter \@firstoftwo
 \else \expandafter \@secondoftwo
 \fi
}%
\providecommand \natexlab [1]{#1}%
\providecommand \enquote  [1]{``#1''}%
\providecommand \bibnamefont  [1]{#1}%
\providecommand \bibfnamefont [1]{#1}%
\providecommand \citenamefont [1]{#1}%
\providecommand \href@noop [0]{\@secondoftwo}%
\providecommand \href [0]{\begingroup \@sanitize@url \@href}%
\providecommand \@href[1]{\@@startlink{#1}\@@href}%
\providecommand \@@href[1]{\endgroup#1\@@endlink}%
\providecommand \@sanitize@url [0]{\catcode `\\12\catcode `\$12\catcode
  `\&12\catcode `\#12\catcode `\^12\catcode `\_12\catcode `\%12\relax}%
\providecommand \@@startlink[1]{}%
\providecommand \@@endlink[0]{}%
\providecommand \url  [0]{\begingroup\@sanitize@url \@url }%
\providecommand \@url [1]{\endgroup\@href {#1}{\urlprefix }}%
\providecommand \urlprefix  [0]{URL }%
\providecommand \Eprint [0]{\href }%
\providecommand \doibase [0]{http://dx.doi.org/}%
\providecommand \selectlanguage [0]{\@gobble}%
\providecommand \bibinfo  [0]{\@secondoftwo}%
\providecommand \bibfield  [0]{\@secondoftwo}%
\providecommand \translation [1]{[#1]}%
\providecommand \BibitemOpen [0]{}%
\providecommand \bibitemStop [0]{}%
\providecommand \bibitemNoStop [0]{.\EOS\space}%
\providecommand \EOS [0]{\spacefactor3000\relax}%
\providecommand \BibitemShut  [1]{\csname bibitem#1\endcsname}%
\let\auto@bib@innerbib\@empty
\bibitem [{\citenamefont {Lieb}\ and\ \citenamefont
  {Robinson}(1972)}]{Lieb1972}%
  \BibitemOpen
  \bibfield  {author} {\bibinfo {author} {\bibfnamefont {E.~H.}\ \bibnamefont
  {Lieb}}\ and\ \bibinfo {author} {\bibfnamefont {D.~W.}\ \bibnamefont
  {Robinson}},\ }\href {\doibase 10.1007/BF01645779} {\bibfield  {journal}
  {\bibinfo  {journal} {Commun. Math. Phys.}\ }\textbf {\bibinfo {volume}
  {28}},\ \bibinfo {pages} {251} (\bibinfo {year} {1972})}\BibitemShut
  {NoStop}%
\bibitem [{\citenamefont {Larkin}\ and\ \citenamefont
  {Ovchinnikov}(1969)}]{Larkin1969}%
  \BibitemOpen
  \bibfield  {author} {\bibinfo {author} {\bibfnamefont {A.~I.}\ \bibnamefont
  {Larkin}}\ and\ \bibinfo {author} {\bibfnamefont {Y.~N.}\ \bibnamefont
  {Ovchinnikov}},\ }\href@noop {} {\bibfield  {journal} {\bibinfo  {journal}
  {Jetp}\ }\textbf {\bibinfo {volume} {28}},\ \bibinfo {pages} {1200} (\bibinfo
  {year} {1969})}\BibitemShut {NoStop}%
\bibitem [{\citenamefont {Jalabert}\ and\ \citenamefont
  {Pastawski}(2001)}]{Jalabert2001a}%
  \BibitemOpen
  \bibfield  {author} {\bibinfo {author} {\bibfnamefont {R.~A.}\ \bibnamefont
  {Jalabert}}\ and\ \bibinfo {author} {\bibfnamefont {H.~M.}\ \bibnamefont
  {Pastawski}},\ }\href {\doibase 10.1103/PhysRevLett.86.2490} {\bibfield
  {journal} {\bibinfo  {journal} {Phys. Rev. Lett.}\ }\textbf {\bibinfo
  {volume} {86}},\ \bibinfo {pages} {2490} (\bibinfo {year}
  {2001})}\BibitemShut {NoStop}%
\bibitem [{\citenamefont {Berman}\ and\ \citenamefont
  {Zaslavsky}(1978)}]{Berman1978}%
  \BibitemOpen
  \bibfield  {author} {\bibinfo {author} {\bibfnamefont {G.}~\bibnamefont
  {Berman}}\ and\ \bibinfo {author} {\bibfnamefont {G.}~\bibnamefont
  {Zaslavsky}},\ }\href {\doibase 10.1016/0378-4371(78)90190-5} {\bibfield
  {journal} {\bibinfo  {journal} {Phys. A Stat. Mech. its Appl.}\ }\textbf
  {\bibinfo {volume} {91}},\ \bibinfo {pages} {450} (\bibinfo {year}
  {1978})}\BibitemShut {NoStop}%
\bibitem [{\citenamefont {Chirikov}\ \emph {et~al.}(1981)\citenamefont
  {Chirikov}, \citenamefont {Izrailev},\ and\ \citenamefont
  {Shepelyansky}}]{Chirikov1981}%
  \BibitemOpen
  \bibfield  {author} {\bibinfo {author} {\bibfnamefont {B.~V.}\ \bibnamefont
  {Chirikov}}, \bibinfo {author} {\bibfnamefont {F.~M.}\ \bibnamefont
  {Izrailev}}, \ and\ \bibinfo {author} {\bibfnamefont {D.~L.}\ \bibnamefont
  {Shepelyansky}},\ }\href@noop {} {\bibfield  {journal} {\bibinfo  {journal}
  {Sov. Sci. Rev.}\ }\textbf {\bibinfo {volume} {2C}},\ \bibinfo {pages} {209}
  (\bibinfo {year} {1981})}\BibitemShut {NoStop}%
\bibitem [{\citenamefont {Chirikov}\ \emph {et~al.}(1988)\citenamefont
  {Chirikov}, \citenamefont {Izrailev},\ and\ \citenamefont
  {Shepelyansky}}]{Chirikov1988}%
  \BibitemOpen
  \bibfield  {author} {\bibinfo {author} {\bibfnamefont {B.}~\bibnamefont
  {Chirikov}}, \bibinfo {author} {\bibfnamefont {F.}~\bibnamefont {Izrailev}},
  \ and\ \bibinfo {author} {\bibfnamefont {D.}~\bibnamefont {Shepelyansky}},\
  }\href {\doibase 10.1016/S0167-2789(98)90011-2} {\bibfield  {journal}
  {\bibinfo  {journal} {Phys. D Nonlinear Phenom.}\ }\textbf {\bibinfo {volume}
  {33}},\ \bibinfo {pages} {77} (\bibinfo {year} {1988})}\BibitemShut {NoStop}%
\bibitem [{\citenamefont {Aleiner}\ and\ \citenamefont
  {Larkin}(1997)}]{Aleiner1997}%
  \BibitemOpen
  \bibfield  {author} {\bibinfo {author} {\bibfnamefont {I.~L.}\ \bibnamefont
  {Aleiner}}\ and\ \bibinfo {author} {\bibfnamefont {A.~I.}\ \bibnamefont
  {Larkin}},\ }\href {\doibase 10.1103/PhysRevE.55.R1243} {\bibfield  {journal}
  {\bibinfo  {journal} {Phys. Rev. E}\ }\textbf {\bibinfo {volume} {55}},\
  \bibinfo {pages} {R1243} (\bibinfo {year} {1997})}\BibitemShut {NoStop}%
\bibitem [{\citenamefont {Sekino}\ and\ \citenamefont
  {Susskind}(2008)}]{sekino_fast_2008}%
  \BibitemOpen
  \bibfield  {author} {\bibinfo {author} {\bibfnamefont {Y.}~\bibnamefont
  {Sekino}}\ and\ \bibinfo {author} {\bibfnamefont {L.}~\bibnamefont
  {Susskind}},\ }\href {\doibase 10.1088/1126-6708/2008/10/065} {\bibfield
  {journal} {\bibinfo  {journal} {J. High Energy Phys.}\ }\textbf {\bibinfo
  {volume} {2008}},\ \bibinfo {pages} {065} (\bibinfo {year}
  {2008})}\BibitemShut {NoStop}%
\bibitem [{\citenamefont {Maldacena}\ \emph {et~al.}(2016)\citenamefont
  {Maldacena}, \citenamefont {Shenker},\ and\ \citenamefont
  {Stanford}}]{Maldacena2016a}%
  \BibitemOpen
  \bibfield  {author} {\bibinfo {author} {\bibfnamefont {J.}~\bibnamefont
  {Maldacena}}, \bibinfo {author} {\bibfnamefont {S.~H.}\ \bibnamefont
  {Shenker}}, \ and\ \bibinfo {author} {\bibfnamefont {D.}~\bibnamefont
  {Stanford}},\ }\href {\doibase 10.1007/JHEP08(2016)106} {\bibfield  {journal}
  {\bibinfo  {journal} {J. High Energy Phys.}\ }\textbf {\bibinfo {volume}
  {2016}},\ \bibinfo {pages} {106} (\bibinfo {year} {2016})}\BibitemShut
  {NoStop}%
\bibitem [{\citenamefont {Kukuljan}\ \emph {et~al.}(2017)\citenamefont
  {Kukuljan}, \citenamefont {Grozdanov},\ and\ \citenamefont
  {Prosen}}]{Kukuljan2017}%
  \BibitemOpen
  \bibfield  {author} {\bibinfo {author} {\bibfnamefont {I.}~\bibnamefont
  {Kukuljan}}, \bibinfo {author} {\bibfnamefont {S.}~\bibnamefont {Grozdanov}},
  \ and\ \bibinfo {author} {\bibfnamefont {T.}~\bibnamefont {Prosen}},\ }\href
  {http://arxiv.org/abs/1701.09147} {\  (\bibinfo {year} {2017})},\ \Eprint
  {http://arxiv.org/abs/1701.09147} {arXiv:1701.09147} \BibitemShut {NoStop}%
\bibitem [{\citenamefont {Kitaev}(2015{\natexlab{a}})}]{Kitaev2015}%
  \BibitemOpen
  \bibfield  {author} {\bibinfo {author} {\bibfnamefont {A.~Y.}\ \bibnamefont
  {Kitaev}},\ }\href {http://online.kitp.ucsb.edu/online/entangled15/kitaev/}
  {\enquote {\bibinfo {title} {{A simple model of quantum holography, Part
  I}},}\ } (\bibinfo {year} {2015}{\natexlab{a}})\BibitemShut {NoStop}%
\bibitem [{\citenamefont {Kitaev}(2015{\natexlab{b}})}]{Kitaev2015a}%
  \BibitemOpen
  \bibfield  {author} {\bibinfo {author} {\bibfnamefont {A.~Y.}\ \bibnamefont
  {Kitaev}},\ }\href {http://online.kitp.ucsb.edu/online/entangled15/kitaev2/}
  {\enquote {\bibinfo {title} {{A simple model of quantum holography, Part
  2}},}\ } (\bibinfo {year} {2015}{\natexlab{b}})\BibitemShut {NoStop}%
\bibitem [{\citenamefont {Georges}\ \emph {et~al.}(2000)\citenamefont
  {Georges}, \citenamefont {Parcollet},\ and\ \citenamefont
  {Sachdev}}]{Georges2000}%
  \BibitemOpen
  \bibfield  {author} {\bibinfo {author} {\bibfnamefont {A.}~\bibnamefont
  {Georges}}, \bibinfo {author} {\bibfnamefont {O.}~\bibnamefont {Parcollet}},
  \ and\ \bibinfo {author} {\bibfnamefont {S.}~\bibnamefont {Sachdev}},\ }\href
  {\doibase 10.1103/PhysRevLett.85.840} {\bibfield  {journal} {\bibinfo
  {journal} {Phys. Rev. Lett.}\ }\textbf {\bibinfo {volume} {85}},\ \bibinfo
  {pages} {840} (\bibinfo {year} {2000})}\BibitemShut {NoStop}%
\bibitem [{\citenamefont {Georges}\ \emph {et~al.}(2001)\citenamefont
  {Georges}, \citenamefont {Parcollet},\ and\ \citenamefont
  {Sachdev}}]{Georges2001}%
  \BibitemOpen
  \bibfield  {author} {\bibinfo {author} {\bibfnamefont {A.}~\bibnamefont
  {Georges}}, \bibinfo {author} {\bibfnamefont {O.}~\bibnamefont {Parcollet}},
  \ and\ \bibinfo {author} {\bibfnamefont {S.}~\bibnamefont {Sachdev}},\ }\href
  {\doibase 10.1103/PhysRevB.63.134406} {\bibfield  {journal} {\bibinfo
  {journal} {Phys. Rev. B}\ }\textbf {\bibinfo {volume} {63}},\ \bibinfo
  {pages} {134406} (\bibinfo {year} {2001})}\BibitemShut {NoStop}%
\bibitem [{\citenamefont {Maldacena}\ and\ \citenamefont
  {Stanford}(2016)}]{Maldacena2016}%
  \BibitemOpen
  \bibfield  {author} {\bibinfo {author} {\bibfnamefont {J.}~\bibnamefont
  {Maldacena}}\ and\ \bibinfo {author} {\bibfnamefont {D.}~\bibnamefont
  {Stanford}},\ }\href {\doibase 10.1103/PhysRevD.94.106002} {\bibfield
  {journal} {\bibinfo  {journal} {Phys. Rev. D}\ }\textbf {\bibinfo {volume}
  {94}},\ \bibinfo {pages} {106002} (\bibinfo {year} {2016})}\BibitemShut
  {NoStop}%
\bibitem [{\citenamefont {Kim}\ and\ \citenamefont {Huse}(2013)}]{Kim2013}%
  \BibitemOpen
  \bibfield  {author} {\bibinfo {author} {\bibfnamefont {H.}~\bibnamefont
  {Kim}}\ and\ \bibinfo {author} {\bibfnamefont {D.~A.}\ \bibnamefont {Huse}},\
  }\href {\doibase 10.1103/PhysRevLett.111.127205} {\bibfield  {journal}
  {\bibinfo  {journal} {Phys. Rev. Lett.}\ }\textbf {\bibinfo {volume} {111}},\
  \bibinfo {pages} {127205} (\bibinfo {year} {2013})}\BibitemShut {NoStop}%
\bibitem [{\citenamefont {Aleiner}\ \emph {et~al.}(2016)\citenamefont
  {Aleiner}, \citenamefont {Faoro},\ and\ \citenamefont {Ioffe}}]{Aleiner2016}%
  \BibitemOpen
  \bibfield  {author} {\bibinfo {author} {\bibfnamefont {I.~L.}\ \bibnamefont
  {Aleiner}}, \bibinfo {author} {\bibfnamefont {L.}~\bibnamefont {Faoro}}, \
  and\ \bibinfo {author} {\bibfnamefont {L.~B.}\ \bibnamefont {Ioffe}},\ }\href
  {http://arxiv.org/abs/1609.01251} {\  (\bibinfo {year} {2016})},\ \Eprint
  {http://arxiv.org/abs/1609.01251} {arXiv:1609.01251} \BibitemShut {NoStop}%
\bibitem [{\citenamefont {Fan}\ \emph {et~al.}(2016)\citenamefont {Fan},
  \citenamefont {Zhang}, \citenamefont {Shen},\ and\ \citenamefont
  {Zhai}}]{Fan2016}%
  \BibitemOpen
  \bibfield  {author} {\bibinfo {author} {\bibfnamefont {R.}~\bibnamefont
  {Fan}}, \bibinfo {author} {\bibfnamefont {P.}~\bibnamefont {Zhang}}, \bibinfo
  {author} {\bibfnamefont {H.}~\bibnamefont {Shen}}, \ and\ \bibinfo {author}
  {\bibfnamefont {H.}~\bibnamefont {Zhai}},\ }\href
  {http://arxiv.org/abs/1608.01914} {\  (\bibinfo {year} {2016})},\ \Eprint
  {http://arxiv.org/abs/1608.01914} {arXiv:1608.01914} \BibitemShut {NoStop}%
\bibitem [{\citenamefont {Basko}\ \emph {et~al.}(2006)\citenamefont {Basko},
  \citenamefont {Aleiner},\ and\ \citenamefont {Altshuler}}]{Basko2006a}%
  \BibitemOpen
  \bibfield  {author} {\bibinfo {author} {\bibfnamefont {D.}~\bibnamefont
  {Basko}}, \bibinfo {author} {\bibfnamefont {I.~L.}\ \bibnamefont {Aleiner}},
  \ and\ \bibinfo {author} {\bibfnamefont {B.~L.}\ \bibnamefont {Altshuler}},\
  }\href {\doibase 10.1016/j.aop.2005.11.014} {\bibfield  {journal} {\bibinfo
  {journal} {Ann. Phys. (N. Y).}\ }\textbf {\bibinfo {volume} {321}},\ \bibinfo
  {pages} {1126} (\bibinfo {year} {2006})}\BibitemShut {NoStop}%
\bibitem [{\citenamefont {Berkelbach}\ and\ \citenamefont
  {Reichman}(2010)}]{Berkelbach2010a}%
  \BibitemOpen
  \bibfield  {author} {\bibinfo {author} {\bibfnamefont {T.~C.}\ \bibnamefont
  {Berkelbach}}\ and\ \bibinfo {author} {\bibfnamefont {D.~R.}\ \bibnamefont
  {Reichman}},\ }\href {\doibase 10.1103/PhysRevB.81.224429} {\bibfield
  {journal} {\bibinfo  {journal} {Phys. Rev. B}\ }\textbf {\bibinfo {volume}
  {81}},\ \bibinfo {pages} {224429} (\bibinfo {year} {2010})}\BibitemShut
  {NoStop}%
\bibitem [{\citenamefont {Luitz}\ \emph {et~al.}(2015)\citenamefont {Luitz},
  \citenamefont {Laflorencie},\ and\ \citenamefont {Alet}}]{Luitz2015}%
  \BibitemOpen
  \bibfield  {author} {\bibinfo {author} {\bibfnamefont {D.~J.}\ \bibnamefont
  {Luitz}}, \bibinfo {author} {\bibfnamefont {N.}~\bibnamefont {Laflorencie}},
  \ and\ \bibinfo {author} {\bibfnamefont {F.}~\bibnamefont {Alet}},\ }\href
  {\doibase 10.1103/PhysRevB.91.081103} {\bibfield  {journal} {\bibinfo
  {journal} {Phys. Rev. B}\ }\textbf {\bibinfo {volume} {91}},\ \bibinfo
  {pages} {081103} (\bibinfo {year} {2015})}\BibitemShut {NoStop}%
\bibitem [{\citenamefont {{\v{Z}}nidari{\v{c}}}\ \emph
  {et~al.}(2008)\citenamefont {{\v{Z}}nidari{\v{c}}}, \citenamefont {Prosen},\
  and\ \citenamefont {Prelov{\v{s}}ek}}]{Znidaric2008}%
  \BibitemOpen
  \bibfield  {author} {\bibinfo {author} {\bibfnamefont {M.}~\bibnamefont
  {{\v{Z}}nidari{\v{c}}}}, \bibinfo {author} {\bibfnamefont {T.}~\bibnamefont
  {Prosen}}, \ and\ \bibinfo {author} {\bibfnamefont {P.}~\bibnamefont
  {Prelov{\v{s}}ek}},\ }\href {\doibase 10.1103/PhysRevB.77.064426} {\bibfield
  {journal} {\bibinfo  {journal} {Phys. Rev. B}\ }\textbf {\bibinfo {volume}
  {77}},\ \bibinfo {pages} {064426} (\bibinfo {year} {2008})}\BibitemShut
  {NoStop}%
\bibitem [{\citenamefont {Bardarson}\ \emph {et~al.}(2012)\citenamefont
  {Bardarson}, \citenamefont {Pollmann},\ and\ \citenamefont
  {Moore}}]{Bardarson2012}%
  \BibitemOpen
  \bibfield  {author} {\bibinfo {author} {\bibfnamefont {J.~H.}\ \bibnamefont
  {Bardarson}}, \bibinfo {author} {\bibfnamefont {F.}~\bibnamefont {Pollmann}},
  \ and\ \bibinfo {author} {\bibfnamefont {J.~E.}\ \bibnamefont {Moore}},\
  }\href {\doibase 10.1103/PhysRevLett.109.017202} {\bibfield  {journal}
  {\bibinfo  {journal} {Phys. Rev. Lett.}\ }\textbf {\bibinfo {volume} {109}},\
  \bibinfo {pages} {017202} (\bibinfo {year} {2012})}\BibitemShut {NoStop}%
\bibitem [{\citenamefont {He}\ and\ \citenamefont {Lu}(2016)}]{He2016}%
  \BibitemOpen
  \bibfield  {author} {\bibinfo {author} {\bibfnamefont {R.-Q.}\ \bibnamefont
  {He}}\ and\ \bibinfo {author} {\bibfnamefont {Z.-Y.}\ \bibnamefont {Lu}},\
  }\href {http://arxiv.org/abs/1608.03586} {\  (\bibinfo {year} {2016})},\
  \Eprint {http://arxiv.org/abs/1608.03586} {arXiv:1608.03586} \BibitemShut
  {NoStop}%
\bibitem [{\citenamefont {Chen}(2016)}]{Chen2016}%
  \BibitemOpen
  \bibfield  {author} {\bibinfo {author} {\bibfnamefont {Y.}~\bibnamefont
  {Chen}},\ }\href {http://arxiv.org/abs/1608.02765} {\  (\bibinfo {year}
  {2016})},\ \Eprint {http://arxiv.org/abs/1608.02765} {arXiv:1608.02765}
  \BibitemShut {NoStop}%
\bibitem [{\citenamefont {Chen}\ \emph {et~al.}(2016)\citenamefont {Chen},
  \citenamefont {Zhou}, \citenamefont {Huse},\ and\ \citenamefont
  {Fradkin}}]{chen_out--time-order_2016}%
  \BibitemOpen
  \bibfield  {author} {\bibinfo {author} {\bibfnamefont {X.}~\bibnamefont
  {Chen}}, \bibinfo {author} {\bibfnamefont {T.}~\bibnamefont {Zhou}}, \bibinfo
  {author} {\bibfnamefont {D.~A.}\ \bibnamefont {Huse}}, \ and\ \bibinfo
  {author} {\bibfnamefont {E.}~\bibnamefont {Fradkin}},\ }\href {\doibase
  10.1002/andp.201600332} {\bibfield  {journal} {\bibinfo  {journal} {Ann.
  Phys. (Berlin)}\ ,\ \bibinfo {pages} {1600332}} (\bibinfo {year}
  {2016})}\BibitemShut {NoStop}%
\bibitem [{\citenamefont {Swingle}\ and\ \citenamefont
  {Chowdhury}(2016)}]{Swingle2016}%
  \BibitemOpen
  \bibfield  {author} {\bibinfo {author} {\bibfnamefont {B.}~\bibnamefont
  {Swingle}}\ and\ \bibinfo {author} {\bibfnamefont {D.}~\bibnamefont
  {Chowdhury}},\ }\href {http://arxiv.org/abs/1608.03280} {\  (\bibinfo {year}
  {2016})},\ \Eprint {http://arxiv.org/abs/1608.03280} {arXiv:1608.03280}
  \BibitemShut {NoStop}%
\bibitem [{\citenamefont {Slagle}\ \emph {et~al.}(2016)\citenamefont {Slagle},
  \citenamefont {Bi}, \citenamefont {You},\ and\ \citenamefont
  {Xu}}]{Slagle2016}%
  \BibitemOpen
  \bibfield  {author} {\bibinfo {author} {\bibfnamefont {K.}~\bibnamefont
  {Slagle}}, \bibinfo {author} {\bibfnamefont {Z.}~\bibnamefont {Bi}}, \bibinfo
  {author} {\bibfnamefont {Y.-Z.}\ \bibnamefont {You}}, \ and\ \bibinfo
  {author} {\bibfnamefont {C.}~\bibnamefont {Xu}},\ }\href
  {http://arxiv.org/abs/1611.04058} {\  (\bibinfo {year} {2016})},\ \Eprint
  {http://arxiv.org/abs/1611.04058} {arXiv:1611.04058} \BibitemShut {NoStop}%
\bibitem [{\citenamefont {Deng}\ \emph {et~al.}(2016)\citenamefont {Deng},
  \citenamefont {Li}, \citenamefont {Pixley}, \citenamefont {Wu},\ and\
  \citenamefont {Sarma}}]{Deng2016a}%
  \BibitemOpen
  \bibfield  {author} {\bibinfo {author} {\bibfnamefont {D.-L.}\ \bibnamefont
  {Deng}}, \bibinfo {author} {\bibfnamefont {X.}~\bibnamefont {Li}}, \bibinfo
  {author} {\bibfnamefont {J.~H.}\ \bibnamefont {Pixley}}, \bibinfo {author}
  {\bibfnamefont {Y.-L.}\ \bibnamefont {Wu}}, \ and\ \bibinfo {author}
  {\bibfnamefont {S.~D.}\ \bibnamefont {Sarma}},\ }\href
  {http://arxiv.org/abs/1607.08611} {\  (\bibinfo {year} {2016})},\ \Eprint
  {http://arxiv.org/abs/1607.08611} {arXiv:1607.08611} \BibitemShut {NoStop}%
\bibitem [{\citenamefont {Hamza}\ \emph {et~al.}(2012)\citenamefont {Hamza},
  \citenamefont {Sims},\ and\ \citenamefont {Stolz}}]{Hamza2012}%
  \BibitemOpen
  \bibfield  {author} {\bibinfo {author} {\bibfnamefont {E.}~\bibnamefont
  {Hamza}}, \bibinfo {author} {\bibfnamefont {R.}~\bibnamefont {Sims}}, \ and\
  \bibinfo {author} {\bibfnamefont {G.}~\bibnamefont {Stolz}},\ }\href
  {\doibase 10.1007/s00220-012-1544-6} {\bibfield  {journal} {\bibinfo
  {journal} {Commun. Math. Phys.}\ }\textbf {\bibinfo {volume} {315}},\
  \bibinfo {pages} {215} (\bibinfo {year} {2012})}\BibitemShut {NoStop}%
\bibitem [{\citenamefont {Friesdorf}\ \emph {et~al.}(2015)\citenamefont
  {Friesdorf}, \citenamefont {Werner}, \citenamefont {Goihl}, \citenamefont
  {Eisert},\ and\ \citenamefont {Brown}}]{Friesdorf}%
  \BibitemOpen
  \bibfield  {author} {\bibinfo {author} {\bibfnamefont {M.}~\bibnamefont
  {Friesdorf}}, \bibinfo {author} {\bibfnamefont {A.~H.}\ \bibnamefont
  {Werner}}, \bibinfo {author} {\bibfnamefont {M.}~\bibnamefont {Goihl}},
  \bibinfo {author} {\bibfnamefont {J.}~\bibnamefont {Eisert}}, \ and\ \bibinfo
  {author} {\bibfnamefont {W.}~\bibnamefont {Brown}},\ }\href {\doibase
  10.1088/1367-2630/17/11/113054} {\bibfield  {journal} {\bibinfo  {journal}
  {New J. Phys.}\ }\textbf {\bibinfo {volume} {17}},\ \bibinfo {pages} {113054}
  (\bibinfo {year} {2015})}\BibitemShut {NoStop}%
\bibitem [{\citenamefont {{Bar Lev}}\ \emph {et~al.}(2015)\citenamefont {{Bar
  Lev}}, \citenamefont {Cohen},\ and\ \citenamefont {Reichman}}]{Lev2014}%
  \BibitemOpen
  \bibfield  {author} {\bibinfo {author} {\bibfnamefont {Y.}~\bibnamefont {{Bar
  Lev}}}, \bibinfo {author} {\bibfnamefont {G.}~\bibnamefont {Cohen}}, \ and\
  \bibinfo {author} {\bibfnamefont {D.~R.}\ \bibnamefont {Reichman}},\ }\href
  {\doibase 10.1103/PhysRevLett.114.100601} {\bibfield  {journal} {\bibinfo
  {journal} {Phys. Rev. Lett.}\ }\textbf {\bibinfo {volume} {114}},\ \bibinfo
  {pages} {100601} (\bibinfo {year} {2015})}\BibitemShut {NoStop}%
\bibitem [{\citenamefont {Agarwal}\ \emph {et~al.}(2015)\citenamefont
  {Agarwal}, \citenamefont {Gopalakrishnan}, \citenamefont {Knap},
  \citenamefont {M{\"{u}}ller},\ and\ \citenamefont {Demler}}]{Agarwal2014}%
  \BibitemOpen
  \bibfield  {author} {\bibinfo {author} {\bibfnamefont {K.}~\bibnamefont
  {Agarwal}}, \bibinfo {author} {\bibfnamefont {S.}~\bibnamefont
  {Gopalakrishnan}}, \bibinfo {author} {\bibfnamefont {M.}~\bibnamefont
  {Knap}}, \bibinfo {author} {\bibfnamefont {M.}~\bibnamefont {M{\"{u}}ller}},
  \ and\ \bibinfo {author} {\bibfnamefont {E.}~\bibnamefont {Demler}},\ }\href
  {\doibase 10.1103/PhysRevLett.114.160401} {\bibfield  {journal} {\bibinfo
  {journal} {Phys. Rev. Lett.}\ }\textbf {\bibinfo {volume} {114}},\ \bibinfo
  {pages} {160401} (\bibinfo {year} {2015})}\BibitemShut {NoStop}%
\bibitem [{\citenamefont {Luitz}\ \emph
  {et~al.}(2016{\natexlab{a}})\citenamefont {Luitz}, \citenamefont
  {Laflorencie},\ and\ \citenamefont {Alet}}]{Luitz2015a}%
  \BibitemOpen
  \bibfield  {author} {\bibinfo {author} {\bibfnamefont {D.~J.}\ \bibnamefont
  {Luitz}}, \bibinfo {author} {\bibfnamefont {N.}~\bibnamefont {Laflorencie}},
  \ and\ \bibinfo {author} {\bibfnamefont {F.}~\bibnamefont {Alet}},\ }\href
  {\doibase 10.1103/PhysRevB.93.060201} {\bibfield  {journal} {\bibinfo
  {journal} {Phys. Rev. B}\ }\textbf {\bibinfo {volume} {93}},\ \bibinfo
  {pages} {060201} (\bibinfo {year} {2016}{\natexlab{a}})}\BibitemShut
  {NoStop}%
\bibitem [{\citenamefont {Gopalakrishnan}\ \emph {et~al.}(2016)\citenamefont
  {Gopalakrishnan}, \citenamefont {Agarwal}, \citenamefont {Demler},
  \citenamefont {Huse},\ and\ \citenamefont {Knap}}]{Gopalakrishnan2015a}%
  \BibitemOpen
  \bibfield  {author} {\bibinfo {author} {\bibfnamefont {S.}~\bibnamefont
  {Gopalakrishnan}}, \bibinfo {author} {\bibfnamefont {K.}~\bibnamefont
  {Agarwal}}, \bibinfo {author} {\bibfnamefont {E.~A.}\ \bibnamefont {Demler}},
  \bibinfo {author} {\bibfnamefont {D.~A.}\ \bibnamefont {Huse}}, \ and\
  \bibinfo {author} {\bibfnamefont {M.}~\bibnamefont {Knap}},\ }\href {\doibase
  10.1103/PhysRevB.93.134206} {\bibfield  {journal} {\bibinfo  {journal} {Phys.
  Rev. B}\ }\textbf {\bibinfo {volume} {93}},\ \bibinfo {pages} {134206}
  (\bibinfo {year} {2016})}\BibitemShut {NoStop}%
\bibitem [{\citenamefont {Luitz}(2016)}]{Luitz2016}%
  \BibitemOpen
  \bibfield  {author} {\bibinfo {author} {\bibfnamefont {D.~J.}\ \bibnamefont
  {Luitz}},\ }\href {\doibase 10.1103/PhysRevB.93.134201} {\bibfield  {journal}
  {\bibinfo  {journal} {Phys. Rev. B}\ }\textbf {\bibinfo {volume} {93}},\
  \bibinfo {pages} {134201} (\bibinfo {year} {2016})}\BibitemShut {NoStop}%
\bibitem [{\citenamefont {Varma}\ \emph {et~al.}(2015)\citenamefont {Varma},
  \citenamefont {Lerose}, \citenamefont {Pietracaprina}, \citenamefont
  {Goold},\ and\ \citenamefont {Scardicchio}}]{Lerose2015}%
  \BibitemOpen
  \bibfield  {author} {\bibinfo {author} {\bibfnamefont {V.~K.}\ \bibnamefont
  {Varma}}, \bibinfo {author} {\bibfnamefont {A.}~\bibnamefont {Lerose}},
  \bibinfo {author} {\bibfnamefont {F.}~\bibnamefont {Pietracaprina}}, \bibinfo
  {author} {\bibfnamefont {J.}~\bibnamefont {Goold}}, \ and\ \bibinfo {author}
  {\bibfnamefont {A.}~\bibnamefont {Scardicchio}},\ }\href
  {http://arxiv.org/abs/1511.09144} {\  (\bibinfo {year} {2015})},\ \Eprint
  {http://arxiv.org/abs/1511.09144} {arXiv:1511.09144} \BibitemShut {NoStop}%
\bibitem [{\citenamefont {{\v{Z}}nidari{\v{c}}}\ \emph
  {et~al.}(2016)\citenamefont {{\v{Z}}nidari{\v{c}}}, \citenamefont
  {Scardicchio},\ and\ \citenamefont {Varma}}]{Znidaric2016}%
  \BibitemOpen
  \bibfield  {author} {\bibinfo {author} {\bibfnamefont {M.}~\bibnamefont
  {{\v{Z}}nidari{\v{c}}}}, \bibinfo {author} {\bibfnamefont {A.}~\bibnamefont
  {Scardicchio}}, \ and\ \bibinfo {author} {\bibfnamefont {V.~K.}\ \bibnamefont
  {Varma}},\ }\href {\doibase 10.1103/PhysRevLett.117.040601} {\bibfield
  {journal} {\bibinfo  {journal} {Phys. Rev. Lett.}\ }\textbf {\bibinfo
  {volume} {117}},\ \bibinfo {pages} {040601} (\bibinfo {year}
  {2016})}\BibitemShut {NoStop}%
\bibitem [{\citenamefont {Khait}\ \emph {et~al.}(2016)\citenamefont {Khait},
  \citenamefont {Gazit}, \citenamefont {Yao},\ and\ \citenamefont
  {Auerbach}}]{Khait2016}%
  \BibitemOpen
  \bibfield  {author} {\bibinfo {author} {\bibfnamefont {I.}~\bibnamefont
  {Khait}}, \bibinfo {author} {\bibfnamefont {S.}~\bibnamefont {Gazit}},
  \bibinfo {author} {\bibfnamefont {N.~Y.}\ \bibnamefont {Yao}}, \ and\
  \bibinfo {author} {\bibfnamefont {A.}~\bibnamefont {Auerbach}},\ }\href
  {\doibase 10.1103/PhysRevB.93.224205} {\bibfield  {journal} {\bibinfo
  {journal} {Phys. Rev. B}\ }\textbf {\bibinfo {volume} {93}},\ \bibinfo
  {pages} {224205} (\bibinfo {year} {2016})}\BibitemShut {NoStop}%
\bibitem [{\citenamefont {Luitz}\ and\ \citenamefont {{Bar
  Lev}}(2016)}]{Luitz2016b}%
  \BibitemOpen
  \bibfield  {author} {\bibinfo {author} {\bibfnamefont {D.~J.}\ \bibnamefont
  {Luitz}}\ and\ \bibinfo {author} {\bibfnamefont {Y.}~\bibnamefont {{Bar
  Lev}}},\ }\href {\doibase 10.1103/PhysRevLett.117.170404} {\bibfield
  {journal} {\bibinfo  {journal} {Phys. Rev. Lett.}\ }\textbf {\bibinfo
  {volume} {117}},\ \bibinfo {pages} {170404} (\bibinfo {year}
  {2016})}\BibitemShut {NoStop}%
\bibitem [{\citenamefont {Luitz}\ and\ \citenamefont
  {Bar~Lev}(2016)}]{luitz_ergodic_2016}%
  \BibitemOpen
  \bibfield  {author} {\bibinfo {author} {\bibfnamefont {D.~J.}\ \bibnamefont
  {Luitz}}\ and\ \bibinfo {author} {\bibfnamefont {Y.}~\bibnamefont
  {Bar~Lev}},\ }\href {http://arxiv.org/abs/1610.08993} {\bibfield  {journal}
  {\bibinfo  {journal} {arXiv:1610.08993}\ } (\bibinfo {year}
  {2016})}\BibitemShut {NoStop}%
\bibitem [{\citenamefont {Damanik}\ \emph {et~al.}(2014)\citenamefont
  {Damanik}, \citenamefont {Lemm}, \citenamefont {Lukic},\ and\ \citenamefont
  {Yessen}}]{Damanik2014}%
  \BibitemOpen
  \bibfield  {author} {\bibinfo {author} {\bibfnamefont {D.}~\bibnamefont
  {Damanik}}, \bibinfo {author} {\bibfnamefont {M.}~\bibnamefont {Lemm}},
  \bibinfo {author} {\bibfnamefont {M.}~\bibnamefont {Lukic}}, \ and\ \bibinfo
  {author} {\bibfnamefont {W.}~\bibnamefont {Yessen}},\ }\href {\doibase
  10.1103/PhysRevLett.113.127202} {\bibfield  {journal} {\bibinfo  {journal}
  {Phys. Rev. Lett.}\ }\textbf {\bibinfo {volume} {113}},\ \bibinfo {pages}
  {127202} (\bibinfo {year} {2014})}\BibitemShut {NoStop}%
\bibitem [{\citenamefont {Bohrdt}\ \emph {et~al.}(2016)\citenamefont {Bohrdt},
  \citenamefont {Mendl}, \citenamefont {Endres},\ and\ \citenamefont
  {Knap}}]{Bohrdt2016}%
  \BibitemOpen
  \bibfield  {author} {\bibinfo {author} {\bibfnamefont {A.}~\bibnamefont
  {Bohrdt}}, \bibinfo {author} {\bibfnamefont {C.~B.}\ \bibnamefont {Mendl}},
  \bibinfo {author} {\bibfnamefont {M.}~\bibnamefont {Endres}}, \ and\ \bibinfo
  {author} {\bibfnamefont {M.}~\bibnamefont {Knap}},\ }\href
  {http://arxiv.org/abs/1612.02434} {\  (\bibinfo {year} {2016})},\ \Eprint
  {http://arxiv.org/abs/1612.02434} {arXiv:1612.02434} \BibitemShut {NoStop}%
\bibitem [{\citenamefont {Nauts}\ and\ \citenamefont
  {Wyatt}(1983)}]{nauts_new_1983}%
  \BibitemOpen
  \bibfield  {author} {\bibinfo {author} {\bibfnamefont {A.}~\bibnamefont
  {Nauts}}\ and\ \bibinfo {author} {\bibfnamefont {R.~E.}\ \bibnamefont
  {Wyatt}},\ }\href {\doibase 10.1103/PhysRevLett.51.2238} {\bibfield
  {journal} {\bibinfo  {journal} {Phys. Rev. Lett.}\ }\textbf {\bibinfo
  {volume} {51}},\ \bibinfo {pages} {2238} (\bibinfo {year}
  {1983})}\BibitemShut {NoStop}%
\bibitem [{\citenamefont {Hernandez}\ \emph {et~al.}(2005)\citenamefont
  {Hernandez}, \citenamefont {Roman},\ and\ \citenamefont
  {Vidal}}]{hernandez_slepc:_2005}%
  \BibitemOpen
  \bibfield  {author} {\bibinfo {author} {\bibfnamefont {V.}~\bibnamefont
  {Hernandez}}, \bibinfo {author} {\bibfnamefont {J.~E.}\ \bibnamefont
  {Roman}}, \ and\ \bibinfo {author} {\bibfnamefont {V.}~\bibnamefont
  {Vidal}},\ }\href {\doibase 10.1145/1089014.1089019} {\bibfield  {journal}
  {\bibinfo  {journal} {ACM Trans. Math. Softw.}\ }\textbf {\bibinfo {volume}
  {31}},\ \bibinfo {pages} {351} (\bibinfo {year} {2005})}\BibitemShut
  {NoStop}%
\bibitem [{\citenamefont {Popescu}\ \emph {et~al.}(2006)\citenamefont
  {Popescu}, \citenamefont {Short},\ and\ \citenamefont
  {Winter}}]{popescu_entanglement_2006}%
  \BibitemOpen
  \bibfield  {author} {\bibinfo {author} {\bibfnamefont {S.}~\bibnamefont
  {Popescu}}, \bibinfo {author} {\bibfnamefont {A.~J.}\ \bibnamefont {Short}},
  \ and\ \bibinfo {author} {\bibfnamefont {A.}~\bibnamefont {Winter}},\ }\href
  {\doibase 10.1038/nphys444} {\bibfield  {journal} {\bibinfo  {journal} {Nat
  Phys}\ }\textbf {\bibinfo {volume} {2}},\ \bibinfo {pages} {754} (\bibinfo
  {year} {2006})}\BibitemShut {NoStop}%
\bibitem [{\citenamefont {Goldstein}\ \emph {et~al.}(2006)\citenamefont
  {Goldstein}, \citenamefont {Lebowitz}, \citenamefont {Tumulka},\ and\
  \citenamefont {Zangh{\`i}}}]{goldstein_canonical_2006}%
  \BibitemOpen
  \bibfield  {author} {\bibinfo {author} {\bibfnamefont {S.}~\bibnamefont
  {Goldstein}}, \bibinfo {author} {\bibfnamefont {J.~L.}\ \bibnamefont
  {Lebowitz}}, \bibinfo {author} {\bibfnamefont {R.}~\bibnamefont {Tumulka}}, \
  and\ \bibinfo {author} {\bibfnamefont {N.}~\bibnamefont {Zangh{\`i}}},\
  }\href {\doibase 10.1103/PhysRevLett.96.050403} {\bibfield  {journal}
  {\bibinfo  {journal} {Phys. Rev. Lett.}\ }\textbf {\bibinfo {volume} {96}},\
  \bibinfo {pages} {050403} (\bibinfo {year} {2006})}\BibitemShut {NoStop}%
\bibitem [{\citenamefont {Reimann}(2007)}]{reimann_typicality_2007}%
  \BibitemOpen
  \bibfield  {author} {\bibinfo {author} {\bibfnamefont {P.}~\bibnamefont
  {Reimann}},\ }\href {\doibase 10.1103/PhysRevLett.99.160404} {\bibfield
  {journal} {\bibinfo  {journal} {Phys. Rev. Lett.}\ }\textbf {\bibinfo
  {volume} {99}},\ \bibinfo {pages} {160404} (\bibinfo {year}
  {2007})}\BibitemShut {NoStop}%
\bibitem [{\citenamefont {Luitz}\ \emph
  {et~al.}(2016{\natexlab{b}})\citenamefont {Luitz}, \citenamefont
  {Laflorencie},\ and\ \citenamefont {Alet}}]{luitz_extended_2016}%
  \BibitemOpen
  \bibfield  {author} {\bibinfo {author} {\bibfnamefont {D.~J.}\ \bibnamefont
  {Luitz}}, \bibinfo {author} {\bibfnamefont {N.}~\bibnamefont {Laflorencie}},
  \ and\ \bibinfo {author} {\bibfnamefont {F.}~\bibnamefont {Alet}},\ }\href
  {\doibase 10.1103/PhysRevB.93.060201} {\bibfield  {journal} {\bibinfo
  {journal} {Phys. Rev. B}\ }\textbf {\bibinfo {volume} {93}},\ \bibinfo
  {pages} {060201} (\bibinfo {year} {2016}{\natexlab{b}})}\BibitemShut
  {NoStop}%
\bibitem [{\citenamefont {Balay}\ \emph {et~al.}(1997)\citenamefont {Balay},
  \citenamefont {Gropp}, \citenamefont {McInnes},\ and\ \citenamefont
  {Smith}}]{petsc-efficient}%
  \BibitemOpen
  \bibfield  {author} {\bibinfo {author} {\bibfnamefont {S.}~\bibnamefont
  {Balay}}, \bibinfo {author} {\bibfnamefont {W.~D.}\ \bibnamefont {Gropp}},
  \bibinfo {author} {\bibfnamefont {L.~C.}\ \bibnamefont {McInnes}}, \ and\
  \bibinfo {author} {\bibfnamefont {B.~F.}\ \bibnamefont {Smith}},\ }in\
  \href@noop {} {\emph {\bibinfo {booktitle} {Modern Software Tools in
  Scientific Computing}}},\ \bibinfo {editor} {edited by\ \bibinfo {editor}
  {\bibfnamefont {E.}~\bibnamefont {Arge}}, \bibinfo {editor} {\bibfnamefont
  {A.~M.}\ \bibnamefont {Bruaset}}, \ and\ \bibinfo {editor} {\bibfnamefont
  {H.~P.}\ \bibnamefont {Langtangen}}}\ (\bibinfo  {publisher}
  {Birkh{\"{a}}user Press},\ \bibinfo {year} {1997})\ pp.\ \bibinfo {pages}
  {163--202}\BibitemShut {NoStop}%
\bibitem [{\citenamefont {Balay}\ \emph
  {et~al.}(2014{\natexlab{a}})\citenamefont {Balay}, \citenamefont {Abhyankar},
  \citenamefont {Adams}, \citenamefont {Brown}, \citenamefont {Brune},
  \citenamefont {Buschelman}, \citenamefont {Eijkhout}, \citenamefont {Gropp},
  \citenamefont {Kaushik}, \citenamefont {Knepley}, \citenamefont {McInnes},
  \citenamefont {Rupp}, \citenamefont {Smith},\ and\ \citenamefont
  {Zhang}}]{petsc-user-ref}%
  \BibitemOpen
  \bibfield  {author} {\bibinfo {author} {\bibfnamefont {S.}~\bibnamefont
  {Balay}}, \bibinfo {author} {\bibfnamefont {S.}~\bibnamefont {Abhyankar}},
  \bibinfo {author} {\bibfnamefont {M.~F.}\ \bibnamefont {Adams}}, \bibinfo
  {author} {\bibfnamefont {J.}~\bibnamefont {Brown}}, \bibinfo {author}
  {\bibfnamefont {P.}~\bibnamefont {Brune}}, \bibinfo {author} {\bibfnamefont
  {K.}~\bibnamefont {Buschelman}}, \bibinfo {author} {\bibfnamefont
  {V.}~\bibnamefont {Eijkhout}}, \bibinfo {author} {\bibfnamefont {W.~D.}\
  \bibnamefont {Gropp}}, \bibinfo {author} {\bibfnamefont {D.}~\bibnamefont
  {Kaushik}}, \bibinfo {author} {\bibfnamefont {M.~G.}\ \bibnamefont
  {Knepley}}, \bibinfo {author} {\bibfnamefont {L.~C.}\ \bibnamefont
  {McInnes}}, \bibinfo {author} {\bibfnamefont {K.}~\bibnamefont {Rupp}},
  \bibinfo {author} {\bibfnamefont {B.~F.}\ \bibnamefont {Smith}}, \ and\
  \bibinfo {author} {\bibfnamefont {H.}~\bibnamefont {Zhang}},\ }\href
  {http://www.mcs.anl.gov/petsc} {\emph {\bibinfo {title} {{PETS}c Users
  Manual}}},\ \bibinfo {type} {Tech. Rep.}\ \bibinfo {number} {ANL-95/11 -
  Revision 3.5}\ (\bibinfo  {institution} {Argonne National Laboratory},\
  \bibinfo {year} {2014})\BibitemShut {NoStop}%
\bibitem [{\citenamefont {Balay}\ \emph
  {et~al.}(2014{\natexlab{b}})\citenamefont {Balay}, \citenamefont {Abhyankar},
  \citenamefont {Adams}, \citenamefont {Brown}, \citenamefont {Brune},
  \citenamefont {Buschelman}, \citenamefont {Eijkhout}, \citenamefont {Gropp},
  \citenamefont {Kaushik}, \citenamefont {Knepley}, \citenamefont {McInnes},
  \citenamefont {Rupp}, \citenamefont {Smith},\ and\ \citenamefont
  {Zhang}}]{petsc-web-page}%
  \BibitemOpen
  \bibfield  {author} {\bibinfo {author} {\bibfnamefont {S.}~\bibnamefont
  {Balay}}, \bibinfo {author} {\bibfnamefont {S.}~\bibnamefont {Abhyankar}},
  \bibinfo {author} {\bibfnamefont {M.~F.}\ \bibnamefont {Adams}}, \bibinfo
  {author} {\bibfnamefont {J.}~\bibnamefont {Brown}}, \bibinfo {author}
  {\bibfnamefont {P.}~\bibnamefont {Brune}}, \bibinfo {author} {\bibfnamefont
  {K.}~\bibnamefont {Buschelman}}, \bibinfo {author} {\bibfnamefont
  {V.}~\bibnamefont {Eijkhout}}, \bibinfo {author} {\bibfnamefont {W.~D.}\
  \bibnamefont {Gropp}}, \bibinfo {author} {\bibfnamefont {D.}~\bibnamefont
  {Kaushik}}, \bibinfo {author} {\bibfnamefont {M.~G.}\ \bibnamefont
  {Knepley}}, \bibinfo {author} {\bibfnamefont {L.~C.}\ \bibnamefont
  {McInnes}}, \bibinfo {author} {\bibfnamefont {K.}~\bibnamefont {Rupp}},
  \bibinfo {author} {\bibfnamefont {B.~F.}\ \bibnamefont {Smith}}, \ and\
  \bibinfo {author} {\bibfnamefont {H.}~\bibnamefont {Zhang}},\ }\href
  {http://www.mcs.anl.gov/petsc} {\enquote {\bibinfo {title} {{PETS}c {W}eb
  page},}\ }\bibinfo {howpublished} {\url{http://www.mcs.anl.gov/petsc}}
  (\bibinfo {year} {2014}{\natexlab{b}})\BibitemShut {NoStop}%
\end{thebibliography}%

\end{document}